\documentclass[twocolumn,showpacs,superscriptaddress,preprintnumbers,amsmath,amssymb]{revtex4}

\usepackage{xspace}
\usepackage{graphicx}
\usepackage{dcolumn}
\usepackage{bm}

\newcommand{\sk}      {Super-Kamiokande\xspace}
\newcommand{\epiz}    {$p \to e^+ \pi^0$\xspace}
\newcommand{\nukp}    {$p \to \bar{\nu} K^+$\xspace}
\newcommand{\pmunu}   {$K^+ \to \mu^+ \nu_{\mu}$\xspace}
\newcommand{\ppipi}   {$K^+ \to \pi^+ \pi^0$\xspace}
\newcommand{\nukz}    {$n \to \bar{\nu} K^0$\xspace}
\newcommand{\mukz}    {$p \to \mu^+ K^0$\xspace}
\newcommand{\ekz}     {$p \to e^+ K^0$\xspace}
\newcommand{\spzpz}   {$K^{0}_{S} \to \pi^0 \pi^0$\xspace}
\newcommand{\spppm}   {$K^{0}_{S} \to \pi^+ \pi^-$\xspace}

\begin{document}

\preprint{APS/123-QED}

\title{Search for nucleon decay via modes favored by supersymmetric grand unification models in Super-Kamiokande-I}

   \newcommand{\icrr}{\affiliation{Kamioka Observatory, Institute for Cosmic Ray Research, University of Tokyo, Kamioka, Gifu, 506-1205, Japan}}
\newcommand{\ncen}{\affiliation{Research Center for Cosmic Neutrinos, Institute for Cosmic Ray Research, University of Tokyo, Kashiwa, Chiba 277-8582, Japan}}
\newcommand{\bu}{\affiliation{Department of Physics, Boston University, Boston, MA 02215, USA}}
\newcommand{\bnl}{\affiliation{Physics Department, Brookhaven National Laboratory, Upton, NY 11973, USA}}
\newcommand{\uci}{\affiliation{Department of Physics and Astronomy, University of California, Irvine, Irvine, CA 92697-4575, USA}}
\newcommand{\csu}{\affiliation{Department of Physics, California State University, Dominguez Hills, Carson, CA 90747, USA}}
\newcommand{\cnu}{\affiliation{Department of Physics, Chonnam National University, Kwangju 500-757, Korea}}
\newcommand{\duke}{\affiliation{Department of Physics, Duke University, Durham, NC 27708 USA}}
\newcommand{\gmu}{\affiliation{Department of Physics, George Mason University, Fairfax, VA 22030, USA}}
\newcommand{\gifu}{\affiliation{Department of Physics, Gifu University, Gifu, Gifu 501-1193, Japan}}
\newcommand{\uh}{\affiliation{Department of Physics and Astronomy, University of Hawaii, Honolulu, HI 96822, USA}}
\newcommand{\ui}{\affiliation{Department of Physics, Indiana University, Bloomington,  IN 47405-7105, USA} }
\newcommand{\kek}{\affiliation{High Energy Accelerator Research Organization (KEK), Tsukuba, Ibaraki 305-0801, Japan}}
\newcommand{\kobe}{\affiliation{Department of Physics, Kobe University, Kobe, Hyogo 657-8501, Japan}}
\newcommand{\kyoto}{\affiliation{Department of Physics, Kyoto University, Kyoto 606-8502, Japan}}
\newcommand{\lanl}{\affiliation{Physics Division, P-23, Los Alamos National Laboratory, Los Alamos, NM 87544, USA}}
\newcommand{\lsu}{\affiliation{Department of Physics and Astronomy, Louisiana State University, Baton Rouge, LA 70803, USA}}
\newcommand{\umd}{\affiliation{Department of Physics, University of Maryland, College Park, MD 20742, USA}}
\newcommand{\duluth}{\affiliation{Department of Physics, University of Minnesota, Duluth, MN 55812-2496, USA}}
\newcommand{\miyagi}{\affiliation{Department of Physics, Miyagi University of Education, Sendai,Miyagi 980-0845, Japan}}
\newcommand{\suny}{\affiliation{Department of Physics and Astronomy, State University of New York, Stony Brook, NY 11794-3800, USA}}
\newcommand{\nagoya}{\affiliation{Department of Physics, Nagoya University, Nagoya, Aichi 464-8602, Japan}}
\newcommand{\niigata}{\affiliation{Department of Physics, Niigata University, Niigata, Niigata 950-2181, Japan}}
\newcommand{\osaka}{\affiliation{Department of Physics, Osaka University, Toyonaka, Osaka 560-0043, Japan}}
\newcommand{\seoul}{\affiliation{Department of Physics, Seoul National University, Seoul 151-742, Korea}}
\newcommand{\shizuokaseika}{\affiliation{International and Cultural Studies, Shizuoka Seika College, Yaizu, Shizuoka 425-8611, Japan}}
\newcommand{\shizuoka}{\affiliation{Department of Systems Engineering, Shizuoka University, Hamamatsu, Shizuoka 432-8561, Japan}}
\newcommand{\skku}{\affiliation{Department of Physics, Sungkyunkwan University, Suwon 440-746, Korea}}
\newcommand{\tohoku}{\affiliation{Research Center for Neutrino Science, Tohoku University, Sendai, Miyagi 980-8578, Japan}}
\newcommand{\tokyo}{\affiliation{University of Tokyo, Tokyo 113-0033, Japan}}
\newcommand{\tokai}{\affiliation{Department of Physics, Tokai University, Hiratsuka, Kanagawa 259-1292, Japan}}
\newcommand{\tit}{\affiliation{Department of Physics, Tokyo Institute for Technology, Meguro, Tokyo 152-8551, Japan}}
\newcommand{\warsaw}{\affiliation{Institute of Experimental Physics, Warsaw University, 00-681 Warsaw, Poland}}
\newcommand{\uw}{\affiliation{Department of Physics, University of Washington, Seattle, WA 98195-1560, USA}}
\newcommand{\tsukubanow}{\altaffiliation{ Present address: Department of Physics, Univ. of Tsukuba, Tsukuba, Ibaraki 305 8577, Japan}}
\newcommand{\okayamanow}{\altaffiliation{ Present address: Department of Physics, Okayama University, Okayama 700-8530, Japan}}
\newcommand{\marylandnow}{\altaffiliation{ Present address: University of Maryland School of Medicine, Baltimore, MD 21201, USA}}
\author{K.Kobayashi}\suny
\author{M.Earl}\marylandnow\bu
%
\author{Y.Ashie}\icrr
\author{J.Hosaka}\icrr
\author{K.Ishihara}\icrr
\author{Y.Itow}\icrr
\author{J.Kameda}\icrr
\author{Y.Koshio}\icrr
\author{A.Minamino}\icrr
\author{C.Mitsuda}\icrr
% ...
\collaboration{The Super-Kamiokande Collaboration}\noaffiliation



\date{\today}

   \begin{abstract}
We report the results for nucleon decay searches via modes favored
by supersymmetric grand unified models in \sk.
Using 1489 days of full \sk -I data, we searched for
\nukp, \nukz, \mukz and \ekz modes.
We found no evidence for nucleon decay in any of these modes.
We set lower limits of partial nucleon lifetime
2.3$\times$10$^{33}$, 1.3$\times$10$^{32}$, 1.3$\times$10$^{33}$ and
1.0$\times$10$^{33}$ years at 90\% confidence level
for \nukp, \nukz, \mukz and \ekz modes, respectively.
These results give a strong constraint on supersymmetric 
grand unification models.
\end{abstract}

\pacs{Valid PACS appear here}
\maketitle

   \section{\label{sec:intro}Introduction} 
Grand Unified Theories (GUTs) \cite{gut1,su5gut} seek to unify the strong and
electroweak forces.  They are motivated by the apparent merging of the
coupling constants of the strong, weak, and electromagnetic forces at a large
energy scale ($\sim10^{16}$ GeV) when low energy measurements are
extrapolated. One of the generic predictions of GUTs is the instability of
the proton, as well as neutrons bound inside the nucleus. The experimental 
observation of nucleon decay would provide a strong evidence of GUTs.

\par
In GUTs, nucleon decay can proceed via an exchange of a massive boson between
two quarks in a proton or in a bound neutron.  In this reaction, one quark
transforms into a lepton and another into an anti-quark which binds with a
spectator quark creating a meson.  The favored decay mode in the prototypical
GUT \cite{su5gut} based on an $SU(5)$ symmetry (``minimal $SU(5)$'')
is $p \to e^+ \pi^0$. For this decay, the proton lifetime scales as $\sim
M_X^4$, where $M_X$ is the mass of the heavy vector gauge boson.  In minimal
$SU(5)$, $M_X$ is on the order of the coupling unification at $10^{15}$
GeV/$c^2$, yielding a predicted proton lifetime of $\tau/B$(\epiz) $\sim
10^{29\pm 2}$ years. The first generation large water Cherenkov detector
experiments \cite{imb,kam}, motivated by this prediction, observed no
evidence of proton decay in this mode and ruled out the model. Also, the
recent result by the \sk experiment extended the previous results
\cite{skepi}. It turns out that this contradiction of $SU(5)$ with the
experimental proton decay limit can be resolved by incorporating
supersymmetry (SUSY) in the theories.

\par
Supersymmetry postulates that, for every SM particle, there is a
corresponding ``superpartner'' with spin differing by 1/2 unit from the SM
particle \cite{susy}. The additional particles stabilize the renormalization
of the Higgs boson and address the so-called ``hierarchy problem''.  When one
incorporates the superpartners into the calculation of the running of the
coupling constants, the convergence of the coupling constants occurs at an
unification scale about one order of magnitude larger than that of minimal
$SU(5)$.  Since the proton decay rate via $p \to e^+ \pi^0$ scales as
$M_X^{-4}$, this leads to a suppression of about four orders of magnitude in
the rate, consistent with experimental non-observation of $p \to e^+ \pi^0$.
Furthermore, while in the minimal $SU(5)$ model, the three coupling constants
do not quite meet at a single point within three standard deviations, they
meet together at a single point in the minimal SUSY $SU(5)$ model
\cite{amaldi}.

\par
However, in many SUSY GUT models, other dominant nucleon decay modes occur via
dimension five operator interactions with the exchange of a heavy
supersymmetric color triplet Higgsino \cite{susysu5}.  These interactions
suppress transitions from one quark family in the initial state to the same
family in the final state.  Since the only second or third generation quark
which is kinematically allowed is the strange quark, an anti-strange quark
typically appears in the final state for these interactions. The anti-strange
quark binds with a spectator quark to form a $K$ meson in the final state.
Thus, SUSY GUTs favor nucleon decays in \nukp and \nukz modes.  The
predictions for the nucleon lifetime in SUSY GUT models, however, varies
widely, and may even be suppressed, since there are many new free parameters
introduced because of the supersymmetry breaking.

\par
In the minimal SUSY $SU(5)$ GUTs, the partial proton lifetime is estimated to
be $\tau/B$(\nukp) $\leq 2.9\times10^{30}$ years \cite{msusysu5}.  Other
models have been proposed beyond those based on $SU(5)$.  In particular,
models based on $SO(10)$ symmetry have become popular in light evidence for 
neutrino mass \cite{skosc}, which they naturally accommodate. An important 
property of the $SO(10)$ symmetry is that there is a heavy right-handed 
neutrino in a multiplet containing the matter fields. 
In addition, all matter fields of one generation can be contained in a single
multiplet, in contrast to theories based on $SU(5)$ where they must be broken
into two separate representations.

\par
One class of models \cite{susymass}, predicts the partial proton lifetime for
\nukp mode to be less than $10^{34}$ years, which is within the observable
range of \sk.  In addition, the same mechanism which gives mass to the
neutrinos provides a new set of dimension five operators through which the
proton can decay.  A consequence of this is that the prediction for \mukz
decay rate is within a factor of 10 of the \nukp decay rate.

\par
These decay modes, favored by SUSY GUT models, have been searched for in
water Cherenkov detector \cite{imb,kam,prenuk} and iron calorimeter
\cite{iron,sou} experiments.  The best limits on the partial nucleon lifetime
via \nukp, \nukz, \mukz, and \ekz are $6.7 \times 10^{32}$ years
\cite{prenuk}, $8.6 \times 10^{31}$ years \cite{kam}, $1.2 \times 10^{31}$
years \cite{kam,imb,sou}, and $1.5 \times 10^{31}$ years \cite{kam},
respectively.

   \section{\label{sec:detec}Super-Kamiokande Detector}
The \sk detector \cite{skdet} is a 50 kton water Cherenkov 
detector located at the Kamioka Observatory of Institute 
for Cosmic Ray Research in the Kamioka mine.
It lies 1,000 m underneath the top of Mt. Ikenoyama,
(i.e. 2,700 m water equivalent underground), resulting in a cosmic ray muon 
rate of 2.2 Hz, a reduction of $10^{-5}$ compared to the rate at the surface.
The detector is optically separated into two regions,
the inner and outer detectors (ID and OD).
The ID of the \sk-I detector, which operated from April 1996 to July 2001,
was instrumented with 11,146 50-cm diameter inward facing 
photomultiplier tubes (PMTs) which provide 40$\%$ photocathode coverage.
This photocathode coverage makes it possible
to detect low energy electrons down to $\sim$5 MeV.
The ID volume is the sensitive region for nucleon decay searches
and has a total fiducial volume of 22.5 kton, defined as a cylindrical volume
with surfaces 2 m away from the ID PMT plane.
The OD completely surrounds the ID
and is instrumented with 1,885 20-cm diameter outward
facing PMTs equipped with 60 cm $\times$ 60 cm wavelength shifter plates to
increase light collection efficiency.
The main purpose of the OD is to tag incoming cosmic ray muons
and exiting muons induced by atmospheric neutrinos.
A detailed description of the \sk-I detector can be found elsewhere \cite{skdet}.

   \section{\label{sec:data}Data Set}
Data used for this analysis is taken during the period
from May 1996 to July 2001. It corresponds to 1489 days
of data taking and an exposure of 92 kt$\cdot$year.

   \section{\label{secl:sim}Nucleon Decay Event and Background Simulation}
In order to search for nucleon decay in the \sk detector, it is critical to 
understand the signal event signature and background characteristics.
We simulate specific decay modes of nucleon decay as well as
background events using a signal event generator and the standard \sk 
atmospheric Monte Carlo (MC). 
By comparing the characteristics of these signal and background simulated 
events in detail, we establish the optimum event selection criteria.
When we limit the nucleon decay events to occur only in the \sk 
fiducial volume, the only significant background to nucleon decays 
originate from atmospheric neutrino interactions. 
Once the selection criteria are established, the detection efficiency is then 
estimated by analyzing the nucleon decay MC sample, and the expected 
background is estimated by analyzing the atmospheric neutrino MC sample.

\subsection{Nucleon decay event simulation}
Nucleon decay in water can occur from a free proton in the hydrogen nucleus
or from a bound nucleon in the oxygen nucleus.
It is relatively simple to simulate the free proton decay using the \sk MC.
However, simulation of the bound nucleon decays requires special care,
because of various nuclear effects experienced by the daughter particles
from nucleon decay before they exit the nucleus. 

\par
\begin{figure}[htbp]
\includegraphics[width=7.5cm]{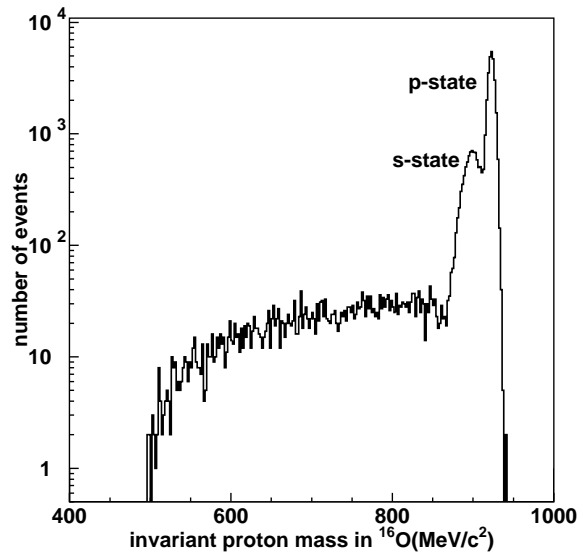}
\caption{\label{fig:sim_protonmass} The invariant proton mass distribution in $^{16}$O in the \nukp MC simulation \cite{yamazaki}. The distributions of s- and p-states are expressed by Gaussian function $G(mean,RMS)$. It is $G(938.3-39.0,10.2))$ MeV/$c^2$ for s-state and $G(938.3-15.5,3.82))$ MeV/$c^2$ for p-state.}
\end{figure}

Nucleons bound in oxygen have Fermi momentum and nuclear binding energy.
In our simulation, we use the Fermi momentum distribution measured by an
electron-$^{12}$C scattering experiment \cite{fermi}.
For nucleon decays in an oxygen nucleus, the nucleon mass must be modified
by nuclear binding energy.
The modified nucleon mass $m_N^{\prime}$ is calculated by $m_N^{\prime}$ =
$m_N$ - $E_{bind}$ where $m_N$ is the nucleon rest mass and $E_{bind}$ is the 
nuclear binding energy.
Yamazaki and Akaishi \cite{yamazaki} estimated the effective nucleon mass
when the nucleon decays in $^{16}$O. Ten percent of decays are from a nucleus
which has wave functions correlated with other nucleons in the nucleus.
Figure \ref{fig:sim_protonmass} shows the invariant proton mass distribution
in $^{16}$O used in the \nukp MC simulation.
The correlated decays produce the broad nucleon mass distribution 
below around 850 MeV/$c^2$.

\par
When a nucleon decays in the oxygen nucleus, the remaining nucleus can be
in an excited state and it can emit prompt gamma-rays during its subsequent
de-excitation process.
This process was studied by Ejiri \cite{ejiri} and we use his results
in this analysis.
This prompt gamma-ray provides us with a powerful tool to tag \nukp events.

\par
The decay products of a nucleon bound in oxygen can interact 
hadronically with the remaining protons and neutrons in the 
residual $^{15}$N nucleus before exiting.  
The initial position of the nucleon is assumed to have the 
Woods-Saxon density distribution \cite{woodssaxon}:
\begin{equation}
\label{nucdens}
\rho(r)=\frac{\rho(0)}{1+\exp{(\frac{r-a}{b})}},
\end{equation}
where $a$ (=1.07 A$^{1/3}$ $=2.69$ fm for $^{16}$O) is the average of nuclear
radius, $2b$ (=0.82 fm) is the thickness of the nuclear surface,
and $r$ is the distance from the center of the nucleus.
Interactions relevant to this analysis are $K^+ N$ and $K^0 N$ elastic 
scatterings, and $K^+ N$ and $K^0 N$ inelastic scatterings via charge exchange.
The $K^+ p$ reactions can proceed elastically or inelastically.
Inelastic interactions are due to the $K^+ p \to K^+ \Delta$.
A partial wave analysis for the scattering amplitudes was performed by
Hyslop {\it et al.} \cite{hyslop} by doing a global fit to many data samples. 
The maximum momentum for the kaon from the nucleon decays is about 600 MeV/$c$.
Below 800 MeV/$c$, $K^+p$ scattering has an extremely small contribution
from inelastic scattering.
Therefore a $K^+$ from a proton decay via \nukp effectively experiences 
only elastic scatterings with protons in the residual nucleus.
Since \nukp are identified through the detection of the daughter 
particles of the kaon decay at rest, $K^+p$ scattering does not affect the 
detection efficiency of \nukp.
The charge exchange reaction $K^+n \to K^0 p$ can reduce the efficiency for
detecting \nukp events.  It is important to estimate what fraction of $K^+$
is lost due to this effect.  The reaction was measured for low momentum
kaons (250 to 600 MeV/$c$) by Glasser {\it et al} \cite{glasser} yielding
cross sections ranging from 2.0 $\pm$ 0.18 mb at $K^+$ momentum of 250 MeV/$c$
to 6.4 $\pm0.56$ mb at $K^+$ momentum of 590 MeV/$c$. To estimate the fraction
of $K^+$ lost due to this effect, a MC simulation is performed.
$K^+$ are started at random points in the nucleus according to the Woods-Saxon
density distribution (Equation \ref{nucdens}).
If there is an interaction, Pauli blocking is taken into account by requiring
the momentum of the recoil nucleon to be above the Fermi surface momentum
($p_F$).
\begin{equation}
\label{pauliblocking}
p_{recoil} \; > \; p_{F}(r) = \hbar \left(\frac{3 \pi^2}{2}\rho(r) \right),
\end{equation}
where $\rho(r)$ is the same as defined in Equation \ref{nucdens}.
From this simulation, it is estimated that $1 \%$ of $K^+$ from \nukp decays
in total are lost because of this charge exchange reaction.
From isospin symmetry, the $K^0 N$ reactions have essentially the
same magnitude as the $K^+ N$ reactions.

\par
We simulate propagation of the produced particles and Cherenkov light in water
by custom code based on GEANT \cite{geant}.
The propagation of charged pions in water is 
simulated by custom code based on \cite{pi16o} for less than 
500 MeV/$c$ and by CALOR \cite{calor} for more than 500 MeV/$c$.

\par
In the \nukz, \mukz and \ekz searches, only decays to $K^0_S$ are studied
because the lifetime of the $K^0_L$ is so long and many of them interact
in water before decaying.
The effect of $K^0$ regeneration is small in $K^0_S$ decay searches.

\par
In total, 50,000 MC events are generated for each decay mode to find
the signature of nucleon decays and to estimate detection efficiencies.
Of these, 34,664, 34,561, 34,648 and 34,572 events are in the fiducial volume
in the \nukp, \nukz, \mukz and \ekz MCs, respectively.

\subsection{Atmospheric neutrino background}
\label{atmnu}
The most significant background to nucleon decay comes from the
atmospheric neutrino interactions in the detector.
Atmospheric neutrinos are produced in  collisions of cosmic rays 
with air molecules in the atmosphere of the earth.
Primary cosmic rays, mostly protons, interact hadronically with air molecules
creating $\pi^\pm$, $K^\pm$, $K^0$, and other mesons.
Neutrinos are then produced from the chain decay of these mesons.
The production of atmospheric neutrinos has been calculated in great detail
by many authors.
For this analysis, we use the calculation by Honda ${\it et}$ ${\it al}$
\cite{atmflx}.

\par
Because the cross section for neutrinos to interact with matter is extremely
small, they can travel unscattered through the earth and interact with a 
nucleon in the water of \sk via the weak interaction.
A generic interaction is
\begin{equation}
\nu_l + N \to l + N^\prime + X,
\end{equation}
where $N$ and $N^\prime$ are the initial and final state nucleons,
$l$ is the outgoing lepton associated with $\nu_l$,
and $X$ can be other possible hadronic particles such as pions.
Because some of these interactions result in topologies similar
to those of nucleon decays, they present a challenging background
to nucleon decay searches. 

\par
We use the same atmospheric neutrino MC simulation \cite{oscfull} that is
used for \sk neutrino oscillation studies with slight modifications.
The modifications made for this analysis are as following:
for single pion production, the so-called axial vector mass which appears
in the neutrino-nucleon differential cross section is set to be 1.2 GeV/$c^2$
to estimate conservatively.
For deep inelastic scattering, we use a model motivated by Bodek and Yang 
\cite{bodek}.

\par
In order to estimate the atmospheric neutrino background events for the 
nucleon decays involving kaons, the production of kaons through the baryon
resonances is included based on the model of Rein and Sehgal \cite{reinsehgal}.
The cross section for kaon production is one or more than one order of 
magnitude smaller than that of single pion production.
In addition, many charged kaons produced in neutrino interactions emit
Cherenkov light while kaons from proton decay do not, and $K$ production
is accompanied by a $\Lambda$ baryon which decays into either $p \pi^-$
or $n \pi^0$ in most cases.
Therefore these background events can be distinguished from nucleon decay
events, and their contributions to the background can be safely ignored.

\par
In order to estimate the backgrounds from the atmospheric neutrinos to
nucleon decays at a high precision, we use a total of 100 year equivalent
sample of atmospheric neutrino MC simulation events for this analysis.

   \section{\label{sec:recon}Event Selection and Reconstruction}
We apply the first stage reduction for the data to remove major background.
Then we reconstruct physical quantities for data and also signal
and background MC events.
Data reduction and event reconstruction is the same as for the neutrino
oscillation analysis \cite{oscfull}.

\subsection{Event Selection}
The trigger threshold for events used in this analysis is set to 29 PMT hits
corresponding to an electron equivalent energy of 5.7 MeV.
The trigger rate at this threshold is about 10 Hz.
The majority of the collected data are the events originating from
cosmic muons and radioactivities.
An event selection process is applied to remove these events.
Almost all cosmic muons are rejected requiring no significant OD activities.
After applying event reconstruction, events are rejected when their electron
equivalence energy is less than 30 MeV corresponding to 197 MeV/c for muons
or when the event vertex is located outside of the fiducial volume.
The remaining event rate after this first stage reduction 
is about eight events/day. 
Almost all of these events originate from atmospheric neutrino interaction.

\subsection{Event Reconstruction}
\begin{figure}[htbp]
\includegraphics[width=7.5cm]{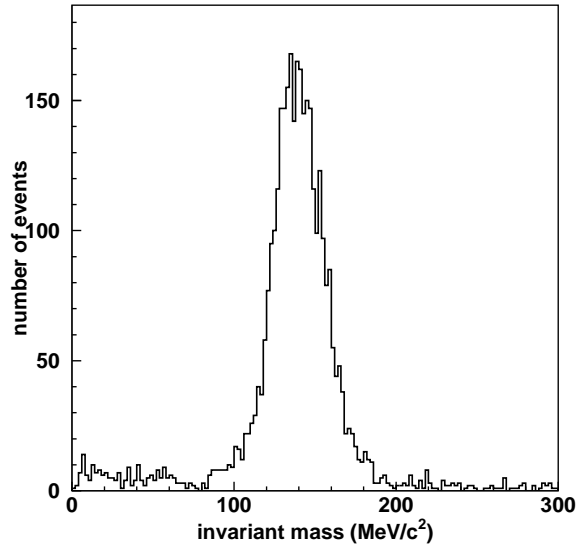}
\caption{\label{fig:pizmass} The reconstructed $\pi^0$ invariant mass distribution in the \nukp,\ppipi MC after applying selection criteria (B1,B2) (See Section \ref{sec:nukp}).}
\end{figure}

Physical quantities are reconstructed for the events remaining
from the event selection process.
In the reconstruction, certain quantities are determined such as the event 
vertex, the number of visible Cherenkov rings, the direction of each ring,
particle identification, momentum and number of Michel electrons.
The vertex resolution is 34 cm for electrons and 25 cm for muons
in single ring events.
In the \nukp, \pmunu and \ppipi MC, the vertex resolution is 47 and 37 cm,
respectively.
In the \nukz, \mukz and \ekz MC, better resolutions are obtained.
Each ring is identified as $e$-like (e$^{\pm}$,$\gamma$) or $\mu$-like 
($\mu^{\pm}$,$\pi^{\pm}$,proton) based on a likelihood analysis of 
Cherenkov ring pattern and Cherenkov angle.
In a single ring electron or muon event, the mis-identification probability
is 0.5\%. Momentum is determined by the sum of photo electrons (PEs) after
correcting for light attenuation in water, PMT angular acceptance, PMT 
coverage, and the assigned electron, muon, or pion particle assumption.
The momentum of a pion is determined by Cherenkov angle as well as the sum of PEs.
The momentum resolution for a 236 MeV/c muon from \nukp,\pmunu is 3\% and
mass resolution for $\pi^0$ from \nukp,\ppipi is 9\%.
Figure \ref{fig:pizmass} shows the reconstructed $\pi^0$ invariant mass
distribution in the \nukp,\ppipi MC.
The detection efficiency of Michel electrons is estimated to be 
80\% for $\mu^+$ and 63\% for $\mu^-$.

\subsection{\label{sec:calib}Calibration}
\par
An important aspect of nucleon decay analyses is the choice of selection
criteria. These criteria are dependent upon the energies
of the particles produced in the various interactions.  
For this reason, the energy deposition and detector response of particles 
traversing the detector must be accurately modeled in the MC simulation.
Energy scale calibrations are performed by comparing the data and MC sample
of Michel electrons from stopping cosmic ray muons, $\pi^0$s produced in
atmospheric neutrino interactions and stopping cosmic ray muons.
The Cherenkov angle and the range are used to obtain muon momentum 
for stopping muons.
These samples provide energy calibration
varying from several tens MeV (Michel electrons)
to several GeV (stopping cosmic ray muons),
entirely covering the energy range relevant to nucleon decay searches.
Based on these samples, the energy scale is reproduced in the MC within 2.5\%.

   \section{\label{sec:ana}Analysis}
During 1,489 days of livetime corresponding to 92 kton$\cdot$year of exposure,
12,179 events are recorded after the first stage reduction process.
Of these, 8,207 are identified as single ring.

\par
It has been well known for some time that there is a deficit of
events induced by atmospheric muon neutrino oscillations \cite{skosc}.
In order to estimate accurately the actual number of background events,
the atmospheric neutrino flux is normalized based on the observed deficit of
$\mu$-like events due to well established neutrino oscillation phenomena and 
the observed number of $e$-like events.
The normalizing factor is determined as multiplying 0.67 for charged current 
$\nu_{\mu}$ events and 1.07 for charged current $\nu_e$ and all neutral
current events.
Finally we normalize the 100 year MC sample (2,246 kton$\cdot$years) to the 
92 kton$\cdot$year sample of \sk exposure.
Based on these simulated signal and background events,
we determine selection criteria for each nucleon decay mode,
which is discussed in the following sections.
All detection efficiencies given below include all relevant branching ratios.

\subsection{\label{sec:nukp}Search for \nukp}
This is the primary mode of proton decay in SUSY GUT models.
A search for \nukp was described in our previous paper \cite{prenuk}
based on the 45 kton$\cdot$year data.
In that paper, we described three methods to search for this mode using the 
two dominant decay modes of the kaon, \ppipi (Method 1) and \pmunu.
For \pmunu, we looked for both the prompt gamma-ray from the de-excitation of
the residual excited $^{15}$N nucleus (Method 2: prompt gamma-ray search) and
for an excess in the momentum distribution of $\mu$-like events at $p_\mu=236$
MeV/$c$ (Method 3: mono-energetic muon search).
In this paper, we present proton lifetime limits derived using these methods
including improved analyses which increase the detection 
efficiency and reduce the background.

\subsubsection{\label{sec:nukpppipi}Method 1: \ppipi search}

If a proton decays to $\bar{\nu}$ K$^+$, the K$^+$ has low enough momentum
that the majority stop before decaying.
Therefore, when the K$^+$ decays to a $\pi^+$ and a $\pi^0$, these two
particles go back-to-back and the $\pi^0$ momentum is expected to be
mono-energetic at 205 MeV/$c$.
To detect this type of events, the following criteria are required:
(A1) two $e$-like rings,
(A2) one Michel electron,
(A3) $175$ MeV/$c < p_{\pi^0} < 250$ MeV/$c$,
(A4) $85$ MeV/$c^2 < m_{\pi^0} < 185$ MeV/$c^2$,
(A5) $40$ PE $< Q_{\pi^+} < 100$ PE, 
(A6)  $Q_{res} < 70$ PE.
The reconstructed total momentum and invariant mass consistent with the
$\pi^0$ from the two $e$-like rings are defined as $p_{\pi^0}$ and 
$m_{\pi^0}$, respectively.
The $\pi^+$ momentum is so close to the Cherenkov threshold 
that the Cherenkov ring is not detected in most cases.
However, since many backgrounds survive criteria (A1-A4),
we use the Cherenkov light produced by the $\pi^+$ in addition.
$Q_{\pi^+}$ is sum of PEs corrected or light attenuation and PMT 
acceptance, which is observed in the PMTs within 40$^{\circ}$ half opening
angle opposite to the $\pi^0$ direction.
$Q_{res}$ is sum of PEs in the remaining PMTs after rejecting the area
within the 90$^{\circ}$ half opening angle toward the two gamma-ray
directions and the $Q_{\pi^+}$ searched area.
We use the criterion (A6) for background rejection.
\begin{figure*}[htbp]
\includegraphics[width=16.5cm]{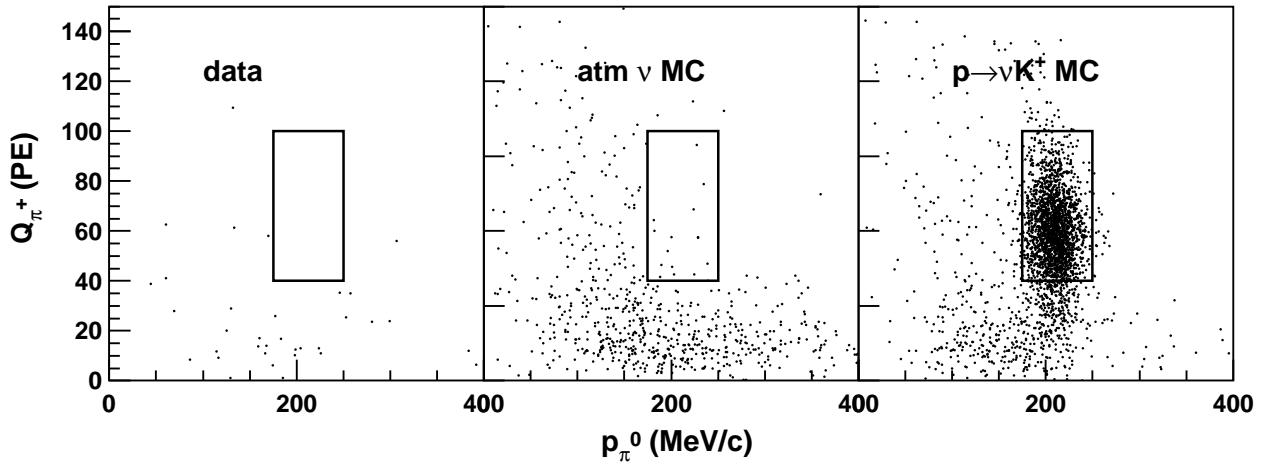}
\caption{\label{fig:nukp_pipi} The distributions of the $Q_{\pi^+}$ versus $p_{\pi^0}$ for events that satisfy criteria (A1,A2,A4,A6). Each figure shows: (left) data, (middle) atmospheric neutrino MC and (right) \nukp MC. The box shows criteria (A3,A5).}
\end{figure*}
The total detection efficiency for this method is estimated to be $6.0\%$
and the background is estimated to be 0.6 events.
Figure \ref{fig:nukp_pipi} shows the distribution of the $Q_{\pi^+}$ versus
$p_{\pi^0}$ for data, the atmospheric neutrino MC and \nukp MC.
The main sources of background are events by single pion production and
deep inelastic scattering.
The incoming neutrino energy of the background
is typically between 0.6 GeV and 2 GeV.
In the data, no events pass the selection criteria (A1-A6).

\subsubsection{\label{sec:nukpgamma}Method 2: \pmunu, prompt gamma-ray search}
When a proton decays in one of the inner shells of the $^{16}$O nucleus,
the residual $^{15}$N nucleus is left in an excited state. This state quickly
de-excites with a certain probability of emitting gamma-rays.
The most probable residual state is a $p_{3/2}$ 6.3 MeV state which leads to an
emission of single 6.3 MeV gamma-ray \cite{ejiri}.
Since the K$^+$ is below the Cherenkov threshold and the K$^+$ lifetime is 
12.4 ns, we can separate the de-excitation gamma-ray from the $\mu^+$ signal.
The $\mu^+$ momentum is mono-energetic (236 MeV/$c$), because the K$^+$ decays
at rest. By requiring a prompt gamma-ray signal as well as the mono-energetic 
muon and an electron from the muon decay, most backgrounds are eliminated.

\par
In order to search for the prompt gamma-ray, three quantities must be defined.
The first is $t_{\mu}$ which is a reference time associated with the detection
of the muon.
The second is $t_0$ which is the time to begin a backward search in time of
flight (TOF) subtracted timing distribution for the earlier hits from the
prompt gamma-ray.
Finally, $t_{\gamma}$ is the time associated with the detection of the gamma-ray.
The reference point $t_{\mu}$ corresponding to the muon is found by searching
for the point in time when $\Delta N_{hit}/\Delta t$ is maximum where
$N_{hit}$ is number of PMT hits.
The starting time $t_0$ is defined as the first point less than $t_{\mu}$
where $dN_{hit}/dt=0$.
In the $t_{\gamma}$ search, PMTs which are within a cone with a $50^\circ$ 
half opening angle with respect to the muon are removed.
Removing these tubes enables the search for the prompt gamma-ray to start at
a closer time to $t_{\mu}$.
A 12 ns timing window is slid backward starting with its trailing edge at $t_0$.
The values $t_{\gamma}$ and $N_{hit\gamma}$ are determined by maximizing
the number of hits in the 12 ns sliding window.
$N_{hit\gamma}$ is the maximum number of hits at $t_{\gamma}$ in the 12 ns
sliding window.

\par
Using these quantities, the following selection criteria are applied:
(B1) one $\mu$-like ring,
(B2) one Michel electron,
(B3) $215$ MeV/$c < p_{\mu} < 260$ MeV/$c$,
(B4) proton rejection,
(B5) $t_{\mu}-t_{\gamma} <$ 100 ns,
(B6) $7 < N_{hit\gamma} < 60$.
Criterion (B4) is applied for rejecting backgrounds caused by poor
vertex reconstruction.
Most of this type of backgrounds is recoil protons produced by neutral
current interactions.
Since the particle type is assumed as a muon in the vertex fit, the 
reconstructed vertex position of a proton is not accurate.
Therefore a fake peak, which mimics gamma-rays, is sometimes produced in the 
TOF subtracted timing distribution.
In order to reject these events, two cuts, $g \geq$ 0.6 and $d_{\mu e} < 200$ cm,
are applied, where $g$ is a goodness of the TOF subtracted timing distribution
and $d_{\mu e}$ is the distance between the Michel positron vertex and the 
muon stopping point.
In proton decay events, $d_{\mu e}$ should be close to zero.

\par
\begin{figure}[htbp]
\includegraphics[width=7.5cm]{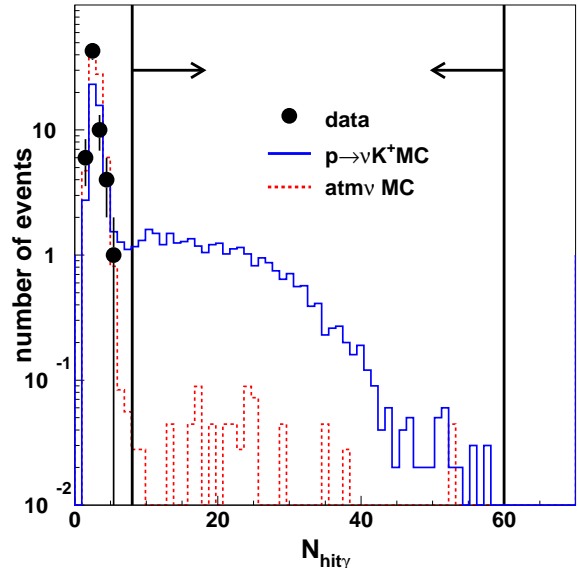}
\caption{\label{fig:nukp_gamma} The $N_{hit\gamma}$ distributions for events that satisfy criteria (B1-B5). The full circles show data and the dashed line shows atmospheric neutrino MC normalized by livetime and neutrino oscillation. The solid line shows \nukp MC. The arrows show criterion (B6), where proton decay candidates with a prompt gamma-ray are expected.}
\end{figure}

Passing the \nukp and 100 year atmospheric neutrino MC events through 
criteria (B1-B6), the detection efficiency and background are estimated
to be 8.6\% and 0.7 events, respectively.
The efficiency includes all branching ratios.
Figure \ref{fig:nukp_gamma} shows $N_{hit\gamma}$ distribution for the 100
year sample of atmospheric neutrino MC normalized by livetime and neutrino
oscillation, \nukp MC, and 92 kton$\cdot$yr sample of data.
No events survive these selection criteria.

\subsubsection{\label{sec:nukpmono}Method 3: \pmunu, mono-energetic muon search}
\begin{figure}[htbp]
\includegraphics[width=7.5cm]{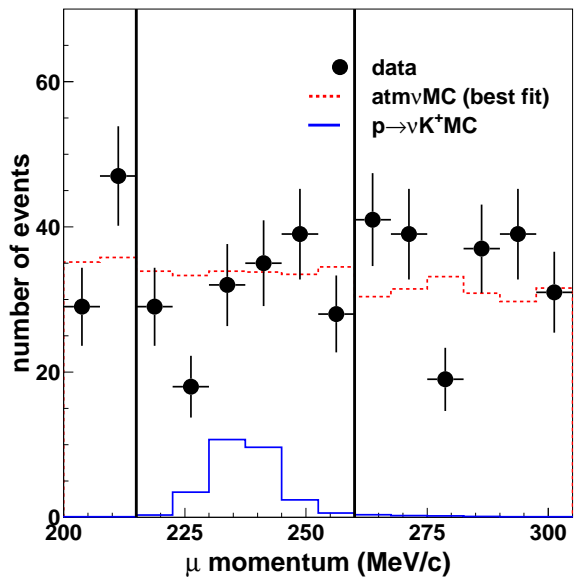}
\caption{\label{fig:nukp_shape} The comparison between data and fitting results of muon momentum distribution for events that satisfy criteria (B1,B2,B7). The full circles show data with statistical errors. The solid line shows \nukp MC. The dashed line shows the best fitted atmospheric neutrino MC with free normalization.}
\end{figure}

We search for a 236 MeV/$c$ mono-energetic muon using events which are not
selected in Method 2. To tag the muon, criteria (B1,B2) are applied, which are 
defined as in Method 2. Moreover, to obtain an independent event sample from 
Method 2, we apply criterion (B7) $N_{hit\gamma} \le 7$.
Figure \ref{fig:nukp_shape} shows the muon momentum distribution of
data compared with best fitted atmospheric neutrino MC.
No significant excess is observed in the signal region (B3).
In this region, there are 181 events with a best fitted 
background of 200 events. 

\subsection{\label{sec:nukz} Search for \nukz}
Many SUSY models also predict the nucleon decay mode, \nukz.
We search for \nukz using the \spzpz and \spppm decay chain.

\par
\begin{figure*}[htbp]
\includegraphics[width=16.5cm]{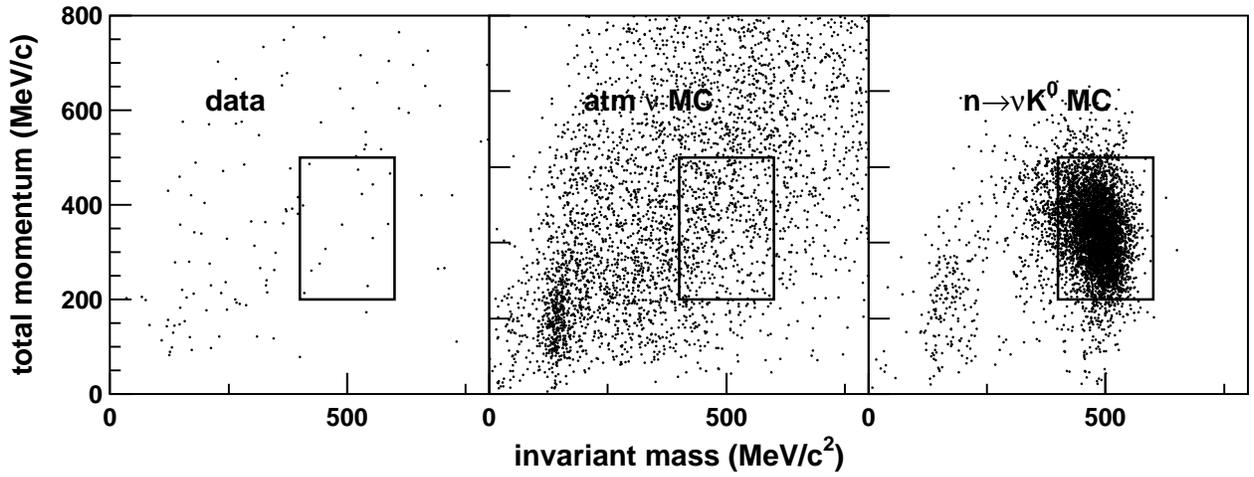}
\caption{\label{fig:nukz_pizpiz} The distributions of the total momentum versus invariant mass for events that satisfy criteria (C1,C2). Each figure shows: (left) data, (middle) atmospheric neutrino MC and (right) \nukz MC. The box shows criteria (C3,C4).}
\end{figure*}

For the \spzpz search, the following criteria are applied:
(C1) three or four $e$-like rings,
(C2) zero Michel electrons,
(C3) 200 MeV/$c < p_{K^0} <$ 500 MeV/$c$,
(C4) 400 MeV/$c^2 < m_{K^0} <$ 600 MeV/$c^2$,
where $p_{K^0}$ and $m_{K^0}$ are total momentum and invariant mass,
assuming the decay sequence of $K^0$.
Because it is difficult to identify four showering Cherenkov rings,
many background events remain. About half of the background events come from
deep inelastic scattering.
Figure \ref{fig:nukz_pizpiz} shows $p_{K^0}$ versus $m_{K^0}$ distributions 
after applying criteria (C1,C2).
The criteria select 14 events in the data with an expected background of
19 and a selection efficiency of 6.9\%.

\par
\begin{figure*}[htbp]
\includegraphics[width=16.5cm]{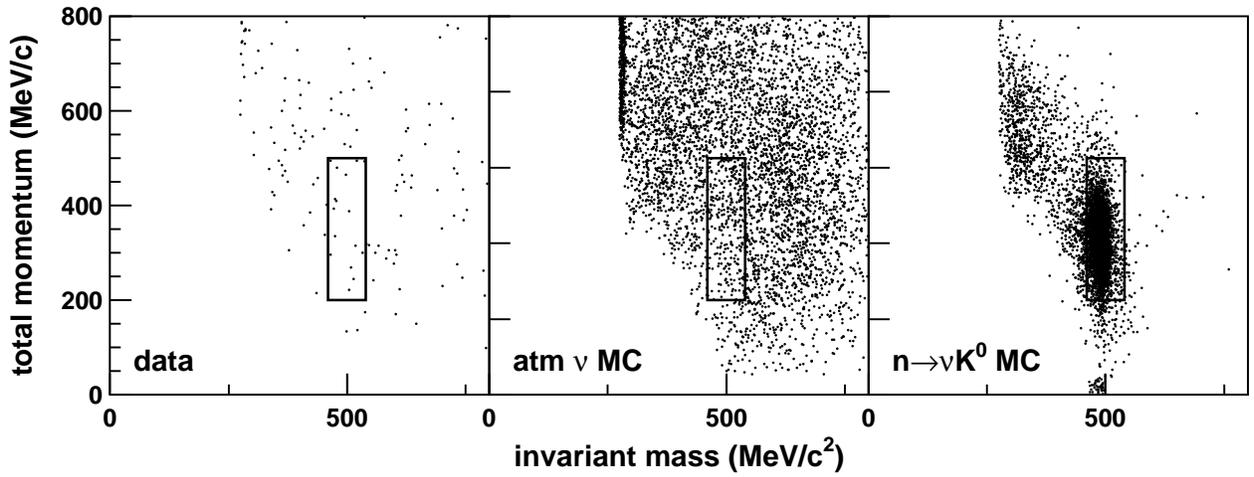}
\caption{\label{fig:nukz_pippim} The distributions of the total momentum versus invariant mass for events that satisfy criteria (D1,D2). Each figure shows: (left) data, (middle) atmospheric neutrino MC and (right) \nukz MC. The box shows criteria (D3,D4). The cluster in the atmospheric neutrino MC figure comes from mis-reconstruction.}
\end{figure*}

For the \spppm search, the following criteria are applied:
(D1) two $\mu$-like rings,
(D2) zero or one Michel electron,
(D3) 200 MeV/$c < p_{K^0} <$ 500 MeV/$c$,
(D4) 450 MeV/$c^2 < m_{K^0} <$ 550 MeV/$c^2$.
Since a large fraction of  $\pi^+$ or $\pi^-$ momenta are below the Cherenkov
threshold, the efficiency of finding both rings is low.
Figure \ref{fig:nukz_pippim} shows $p_{K^0}$ versus $m_{K^0}$ distributions 
after applying criteria (D1,D2).
Twenty events are observed in the data with the detection efficiency of 5.5\%,
while 11.2 background events are expected.
Most of the background events are produced by charged current single pion
production.


\subsection{\label{sec:mukz}Search for \mukz}
In a SUSY $SO(10)$ model \cite{susymass} with neutrino mass, \mukz is an
important decay mode.
We search for \mukz using \spzpz and \spppm decay chain.
In this mode, the total invariant proton mass and momentum can be reconstructed
and the backgrounds can be significantly reduced.

\par
For the \spzpz search, the following criteria are applied:
(E1) 2-4 $e$-like rings and one $\mu$-like ring,
(E2) zero or one Michel electron,
(E3) 400 MeV/$c^2 < m_{K^0} <$ 600 MeV/$c^2$,
(E4) 150 MeV/$c < p_{\mu} <$ 400 MeV/$c$,
(E5) $p_{p} <$ 300 MeV/$c$,
(E6) 750 MeV/$c^2 < m_{p} <$ 1000 MeV/$c^2$,
where $m_{p}$ and $p_{p}$ are the invariant mass and total momentum 
assuming the decay sequence of proton, respectively.
The momentum of the $\mu^+$, $p_{\mu}$, is determined using the $\mu$-like ring.
Figure \ref{fig:mukz_pizpiz} shows $p_p$ versus $m_p$ distributions 
after applying criteria (E1-E4).
No events survive in the data with the estimated 5.4\% detection efficiency
and 0.4 events of the expected background.

\par
\begin{figure*}[htbp]
\includegraphics[width=16.5cm]{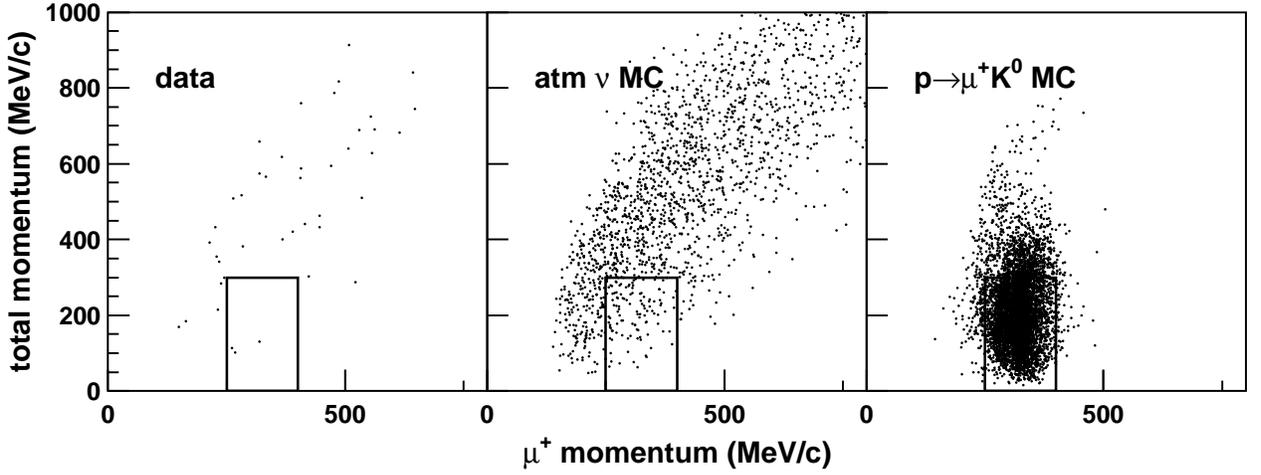}
\caption{\label{fig:mukz_pippim1} The distributions of the total momentum versus $\mu^+$ momentum for events that satisfy criteria (F1,F2). Each figure shows: (left) data, (middle) atmospheric neutrino MC and (right) \mukz MC. The box shows criteria (F3,F4).}
\end{figure*}

For the \spppm search, two different methods are applied,
because ring finding efficiency of the two pions are low.
The first method (Method 1) uses the following criteria:
(F1) two $\mu$-like rings,
(F2) two Michel electrons,
(F3) 250 MeV/$c < p_{\mu} <$ 400 MeV/$c$,
(F4) $p_{p} <$ 300 MeV/$c$.
We assume the more energetic ring as muon and the other as charged pion.
Figure \ref{fig:mukz_pippim1} shows $p_p$ versus $p_{\mu}$ distributions 
after applying criteria (F1,F2).
The detection efficiency and the expected background are
7.0\% and 3.2 events, respectively; 3 events are observed in the data.
The second method (Method 2) requires:
(G1) three rings,
(G2) one or two Michel electrons,
(G3) 450 MeV/$c^2 < m_{K^0} <$ 550 MeV/$c^2$,
(G4) $p_{p} <$ 300 MeV/$c$,
(G5) 750 MeV/$c^2 < m_{p} <$ 1000 MeV/$c^2$.
Figure \ref{fig:mukz_pippim2} shows the $p_p$ versus $m_p$ distributions 
after applying criteria (G1-G3).
The detection efficiency and the expected background are estimated to be
2.8\% and 0.3 events, respectively.
Most of the background events come from charged current $\nu_{\mu}$ single pion
production. No surviving events are observed in the data.

\begin{figure*}[htbp]
\includegraphics[width=16.5cm]{./fig/mukz_pizpiz.epsi}
\caption{\label{fig:mukz_pizpiz} The distributions of the total momentum versus invariant mass for events that satisfy criteria (E1-E4). Each figure shows: (left) data, (middle) atmospheric neutrino MC and (right) \mukz MC. The box shows criteria (E5,E6).}
\end{figure*}
\begin{figure*}[htbp]
\includegraphics[width=16.5cm]{./fig/mukz_pippim2.epsi}
\caption{\label{fig:mukz_pippim2} The distributions of the total momentum versus invariant mass for events that satisfy criteria (G1-G3). Each figure shows: (left) data, (middle) atmospheric neutrino MC and (right) \mukz MC. The box shows criteria (G4,G5).}
\end{figure*}

\subsection{\label{sec:ekz}Search for \ekz}
One of the supersymmetric theories, based on flavor group $(S_3)^3$
predicts that \ekz occurs at a comparable rate with \nukp and \nukz \cite{s3}.
For the search for \ekz mode, three methods are applied in the same way
as \mukz search.
For the \spzpz search, the selection criteria are:
(H1) 3-5 $e$-like rings,
(H2) zero Michel electrons,
(H3) $p_{p} <$ 300 MeV/$c$,
(H4) 750 MeV/$c^2 < m_{p} <$ 1000 MeV/$c^2$.
We assume the most energetic ring as electron and others as gamma-rays from
the $\pi^0$ decays.
Figure \ref{fig:ekz_pizpiz} shows $p_p$ versus $m_p$ distributions 
after applying criteria (H1,H2).
The detection efficiency and the expected background are
9.2\% and 1.1 events, respectively.
One candidate event is observed in the data.

\par
\begin{figure*}[htbp]
\includegraphics[width=16.5cm]{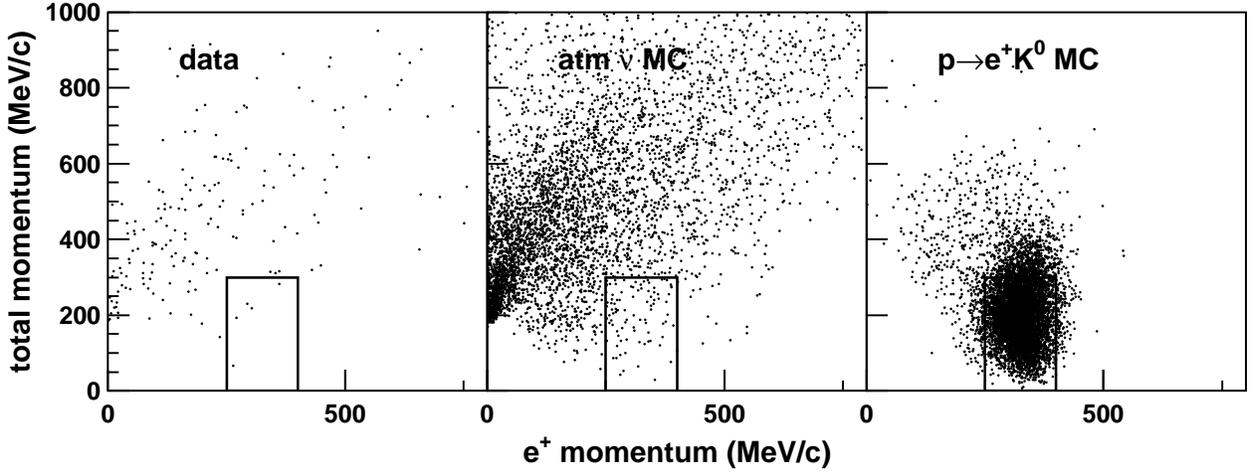}
\caption{\label{fig:ekz_pippim1} The distributions of the total momentum versus $e^+$ momentum for events that satisfy criteria (I1,I2). Each figure shows: (left) data, (middle) atmospheric neutrino MC and (right) \ekz MC. The box shows criteria (I3,I4).}
\end{figure*}

For the \spppm search, two different methods are applied.
The first method (Method 1) uses the following criteria:
(I1) one $\mu$-like ring and one $e$-like ring,
(I2) one Michel electron,
(I3) 250 MeV/$c < p_e<$ 400 MeV/$c$,
(I4) $p_{p} <$ 300 MeV/$c$,
where $p_e$ is the electron momentum determined using the $e$-like ring.
Figure \ref{fig:ekz_pippim1} shows $p_p$ versus $p_e$ distributions 
after applying criteria (I1,I2).
The detection efficiency and the expected background are
7.9\% and 3.6 events, respectively.
Most of the background events are produced by charged current $\nu_e$ single
pion events. Five events are observed in the data.
The second method (Method 2) uses the criteria:
(J1) three rings (at least one $e$-like),
(J2) zero or one Michel electron,
(J3) 450 MeV/$c^2 < m_{K^0} <$ 550 MeV/$c^2$,
(J4) $p_{p} <$ 300 MeV/$c$,
(J5) 750 MeV/$c^2 < m_{p} <$ 1000 MeV/$c^2$.
Figure \ref{fig:ekz_pippim2} shows $p_p$ versus $m_p$ distributions 
after applying criteria (J1-J3).
The detection efficiency is estimated to be 1.3\% with 0.04 
expected backgrounds. No surviving events are observed in the data.

\begin{figure*}[htbp]
\includegraphics[width=16.5cm]{./fig/ekz_pizpiz.epsi}
\caption{\label{fig:ekz_pizpiz} The distributions of the total momentum versus invariant mass for events that satisfy criteria (H1,H2). Each figure shows: (left) data, (middle) atmospheric neutrino MC and (right) \ekz MC. The box shows criteria (H3,H4).}
\end{figure*}
\begin{figure*}[htbp]
\includegraphics[width=16.5cm]{./fig/ekz_pippim2.epsi}
\caption{\label{fig:ekz_pippim2} The distributions of the total momentum versus invariant mass for events that satisfy criteria (J1-J3). Each figure shows: (left) data, (middle) atmospheric neutrino MC and (right) \ekz MC. The box shows criteria (J4,J5).}
\end{figure*}

\subsection{\label{sec:sys}Systematic uncertainties}
In the detection efficiency, we consider the following common systematic
uncertainties in every search: imperfect knowledge of light scattering 
in water, energy scale, and particle identification.
The uncertainty of the rate of light scattering in water is estimated
to be 20\%, which mainly effects on the ring finding efficiency.
The energy scale uncertainty is estimated to be 2.5\% by the calibration
described in Section \ref{sec:calib}.
For the modes which have a charged pion from kaon decay, we consider the
imperfect knowledge of a charged pion-nucleon cross section in water which
is estimated to be 10\% by comparing with our MC and experimental data 
\cite{piocrs}.
For the \mukz and \ekz search, we consider the uncertainty of the Fermi 
momentum which is estimated to be about 5\% from model differences \cite{fermi}.
In addition, we consider the uncertainty of de-excitation gamma-ray emission
probabilities in the \nukp, prompt gamma-ray search. 
It is estimated to be 15\% for 6.3 MeV gamma-ray and 30\% for 
other gamma-rays \cite{ejiri}.
From these sources of systematic uncertainty,
the total contribution to the uncertainty of the detection efficiency
in the \nukp, prompt gamma-ray search is estimated to be 20\%.
The main uncertainty comes from the uncertainty of de-excitation
gamma-ray emission probabilities.
The total contributions in the \nukp, mono-energetic and \ppipi search are
estimated to be 2.5\% and 8.8\%, respectively.
Most of the uncertainties in the \ppipi search come from the imperfect
knowledge of charged pion-nucleon cross section in water and light
scattering in water, which are estimated to be 4.8\% and 6.7\%, respectively.
In the \nukz search, the total contributions are
estimated to be 16\% and 14\% for the \spzpz and \spppm search, respectively.
The main source of uncertainty comes from the uncertainty in
light scattering in water. 
In all methods of the \mukz and \ekz searches, the total contributions
are less than 20\%.

\par
In the background estimation, we consider the following common systematic
uncertainties in all searches:
imperfect knowledge of atmospheric neutrino flux, neutrino cross sections,
energy scale, and particle identification.
Because we use the scaled MC, the absolute normalization error becomes smaller.
However, the background uncertainty from the atmospheric neutrino flux 
normalization is conservatively estimated to be 20\% \cite{atmflx}.
The imperfect knowledge of the cross sections are considered to be 30\%
for quasi-elastic and elastic scattering, 30\% for single meson production
and 50\% for deep inelastic scattering.
From these sources of systematic uncertainty, the total contributions to the 
uncertainty of the background in the \nukz, \spzpz and \spppm search are 
estimated to be 44\% and 41\%, respectively.
The contributions in many other modes become more than 50\%, because
background statistics of the atmospheric neutrino MC are poor.
The uncertainties in all of the searches are summarized 
at Table \ref{tab:res-sum}.

\subsection{\label{sec:result}Results}
In the absence of any significant nuleon decay signature, we interpret
our results as lower limits of nucleon partial lifetime for each decay mode
using the following method \cite{limitcalc}.

\par
First, based on Bayes theorem, we calculate the nucleon decay probability, 
$P(\Gamma|n_{i})$, as follows:
\begin{widetext}
\begin{equation}
\label{eq:prob}
P(\Gamma|n_{i})=A\int\!\int\!\int\frac{e^{-(\Gamma\lambda_i\epsilon_i+b_i)}(\Gamma\lambda_i\epsilon_i+b_i)^{n_i}}{n_i!}P(\Gamma)P(\lambda_i)P(\epsilon_i)P(b_i)d\lambda_id\epsilon_idb_i.
\end{equation}
\end{widetext}
Here a Poisson distribution is assumed for the nucleon decay probability;
$n_i$ is the number of candidate events in the $i$-th nucleon decay search;
$\Gamma$ is the total decay rate;
$\lambda_i$ is the corresponding detector exposure;
$\epsilon_i$ is the detection efficiency including the meson branching ratio;
and $b_i$ is the expected background.
In our search, $\lambda_i$ is 3.05$\times 10^{34}$ proton$\cdot$year 
for proton decay and 2.44$\times 10^{34}$ neutron$\cdot$year for neutron decay.
$P(\Gamma)$ is the decay rate probability density. We assume $P(\Gamma)$
as one for $\Gamma>0$ and otherwise zero.
The uncertainties of detector exposure ($P(\lambda_i)$), detection efficiency 
($P(\epsilon_i)$) and background ($P(b_i)$) are expressed as follows:
\begin{equation}
\label{eq:densexp}
P(\lambda_i) = \delta(\lambda_i - \lambda_{0,i})
\end{equation}
\begin{equation}
\label{eq:denseff}
P(\epsilon_i) = e^{-(\epsilon_i - \epsilon_{0,i})^2/2\sigma_{\epsilon,i}^2} \: (0\leq \epsilon_i \leq 1, \: {\rm otherwise} \: 0)
\end{equation}
\begin{eqnarray}
\label{eq:densbkg}
& & P(b_i) = \frac{1}{b_i}\int_{0}^{\infty}\frac{e^{-b^{\prime}}(b^{\prime})^{n_{b,i}}}{n_{b,i}!}e^{-\frac{-(b_iC_i-b^{\prime})^2}{2\sigma_{b,i}^2}}db^{\prime} \nonumber\\
& & \hspace{3.cm} (0\leq b_i, \: {\rm otherwise} \: 0),
\end{eqnarray}
where $\lambda_{0,i}$ is the estimated exposure, 
$\epsilon_{0,i}$ is the estimated detection efficiency,
$\sigma_{\epsilon,i}$ is the estimated uncertainty in the detection efficiency,
$n_{b,i}$ is the number of background events,
$C_i$ is the MC oversampling factor, 
and $\sigma_{b,i}$ is the uncertainty in the background.
Because the uncertainty in the 
exposure is small, a delta function is assumed for $P(\lambda_i)$.
The lower limit of the nucleon decay rate, $\Gamma_{limit}$, is calculated using
Equation \ref{eq:cl}, where $n$ is the number of searches for a decay mode.
In our search, we calculate 90\% confidence level (CL) limit, i.e., $CL = 0.9$
as follows:
\begin{equation}
\label{eq:cl}
CL = \frac{\displaystyle{ \int_{\Gamma=0}^{\Gamma_{limit}}\prod_{i=1}^nP(\Gamma|n_i)d\Gamma}}{\displaystyle{\int_{\Gamma=0}^{\infty}\prod_{i=1}^nP(\Gamma|n_i)d\Gamma}}.
\end{equation}
The lower limit of partial nucleon lifetime, $\tau/B$,
is now calculated by:
\begin{equation}
\label{eq:limit}
\tau/B = \frac{1}{\Gamma_{limit}}\sum_{i=1}^{n} [\epsilon_{0,i}\cdot \lambda_{0,i}].
\end{equation}

\par
In the \nukp, prompt gamma-ray and \ppipi search, the lower limit of 
the nucleon partial lifetime is found to be $1.0\times 10^{33}$ years and
$7.8\times 10^{32}$ years at 90\% CL, respectively.
Only for \nukp, mono-energetic muon search, we use a special method because it
has a lot of background. We search for an excess in the signal region (B3)
of the muon momentum distribution.
After applying criteria (B1,B2,B7), the events are divided into three momentum
bins; $p_{\mu}$ are 200-215 MeV/$c$, 215-260 MeV/$c$ and 260-305 MeV/$c$.
The numbers of events in each momentum bin ($n_1$, $n_2$, $n_3$) are 76, 181,
185 events, respectively.
The expected numbers of neutrino MC events in each momentum bin 
($b_{1}$, $b_{2}$, $b_{3}$) are 78, 223, 182 events, respectively.
The nucleon decay probabilities, $P(\Gamma|n_1, n_2, n_3)$, is then calculated
using Equation \ref{eq:shape}:
\begin{widetext}
\begin{equation}
\label{eq:shape}
P(\Gamma|n_1, n_2, n_3)=A\int\!\int\!\int\prod_{i=1}^{3}\frac{e^{-(\Gamma\lambda_i\epsilon_i+b_{shape,i}*b)}(\Gamma\lambda_i\epsilon_i+b_{shape,i}*b)^{n_i}}{n_i!}P(\Gamma)P(\lambda_i)P(\epsilon_i)P(b)P(b_{shape,i})d\lambda_id\epsilon_idbdb_{shape,i},
\end{equation}
\end{widetext}
where $i=1,2,3$, corresponds to 200-215 MeV/$c$, 215-260 MeV/$c$ and
260-305 MeV/$c$, respectively; 
$P(b)$ is defined as one for $0<b$ and otherwise zero; $\epsilon_1$, 
$\epsilon_2$, and $\epsilon_3$ are estimated to be 0.25\%, 34\% and 1.3\%,
respectively.
The background shape, $b_{shape,i}$, is $b_i$ divided by $b_2$.
The uncertainty function of the background shape $P(b_{shape,i})$ is defined 
to be a Gaussian function for $i=1,3$ and a delta function for $i=2$.
The uncertainties for $i=1,3$ are then estimated to be 7\% and 8\% from the MC
model difference, respectively.
From Equations \ref{eq:cl} and \ref{eq:limit} using
$P(\Gamma|n_1, n_2, n_3)$ instead of $P(\Gamma|n_{i})$, decay limit
is calculated to be $6.4\times10^{32}$ years at 90$\%$ CL.
Using the three methods, the combined lower limit on the partial lifetime
of proton via \nukp is $2.3 \times 10^{33}$ years at 90\% CL.

\par
In the \nukz analysis, although we observe more events than the expected
background in the \spppm search, there is no excess in the \spzpz search. 
Therefore we also set a nucleon decay lifetime limit in this article. 
Combining two methods, we obtain the lifetime limit of 
$1.3\times10^{32}$ years at 90\% CL.
In the \mukz and \ekz search, by combining three methods, the lifetime lower
limits are set to be $1.3\times10^{33}$ years and $1.0\times10^{33}$ at 
90\% CL, respectively.
The limits in all searches are summarized at Table \ref{tab:res-sum}.
\begin{table*}
\caption{\label{tab:res-sum} Summary of nucleon decay search. The numbers in the parentheses are the systematic uncertainties (\%). }
\begin{center}
\begin{tabular}{|cl|r|r|r|r|}
\hline
\hline
mode & method & efficiency & background & candidate & lower limit\\
& & (\%) &  & & ($\times$10$^{32}$years)\\
\hline
\hline
\nukp & total & & & & 23 \\
 & prompt gamma-ray search    &  8.6 (20.) & 0.7 (59.) & 0   & 10. \\
 & mono-energetic muon search & 35.6 (2.5) & --- & --- &  6.4 \\
 & \ppipi search   &  6.0 (8.8) & 0.6 (74.) & 0   &  7.8 \\
\hline
\nukz & total & & & & 1.3 \\
 & \spzpz & 6.9 (16.) & 19. (44.) & 14 & 1.3 \\
 & \spppm & 5.5 (14.) & 11. (41.)  & 20 & 0.69 \\
\hline
\mukz & total & & & & 13 \\
 & \spzpz        & 5.4 (11.) & 0.4 (78.) & 0 & 7.0 \\
 & \spppm Method 1 & 7.0 (9.5) & 3.2 (41.) & 3 & 4.4 \\
 & \spppm Method 2 & 2.8 (12.) & 0.3 (76.) & 0 & 3.6 \\
\hline
\ekz & total & & & & 10 \\
 & \spzpz        & 9.2 (5.8) & 1.1 (62.) & 1 & 8.4 \\
 & \spppm Method 1 &  7.9 (12.) & 3.6 (50.) & 5 & 3.5 \\
 & \spppm Method 2 &  1.3 (19.) & 0.04 (146.) & 0 & 1.6 \\
\hline
\hline
\end{tabular}
\end{center}
\end{table*}

   \section{\label{sec:conc}Conclusion}
We have searched for nucleon decay via \nukp, \nukz, \mukz and \ekz
from an exposure of 92 kt$\cdot$year.
No significant excess above background is observed.
The lower limits of the partial nucleon lifetime
at 90\% CL for each mode are $2.3\times10^{33}$, $1.3\times10^{32}$,
$1.3\times10^{33}$ and $1.0\times10^{33}$ years, respectively
(See Table \ref{tab:res-sum}).
From these results minimal SUSY SU(5) is fully excluded \cite{msusysu5}.
These results also give strong constraints on other SUSY GUT models 
\cite{susysu5,susymass}.

\par
We gratefully acknowledge the cooperation of the Kamioka Mining and
Smelting Company.  The \sk experiment has been built and
operated from funding by the Japanese Ministry of Education, Culture,
Sports, Science and Technology, the United States Department of Energy,
and the U.S. National Science Foundation.
Some of us have been supported by funds from the Korean Research
Foundation (BK21) and the Korea Science and Engineering Foundation, 
the Polish Committee for Scientific Research (grant 1P03B08227),
Japan Society for the Promotion of Science, and
Research Corporation's Cottrell College Science Award.


\end{document}